%% file: BHterms.tex
\documentclass[12pt,a4paper]{article}

\usepackage{cite}
\usepackage{amsmath}
\usepackage{amssymb}
\usepackage{graphicx}

\lccode`\-=`\-

\makeatletter

\AtBeginDocument{
   \let\@@to=\to \def\to{\@@to\nobreak\discretionary{}{\hbox{$\@@to$}}{}}
}

\renewcommand{\section}{\@startsection{section}{1}
	{0ex}{3.5ex plus 1ex minus .2ex}{2.3ex plus .2ex}
	{\normalfont\bfseries}}
\def\@seccntformat#1{\csname the#1\endcsname.~}

\long\def\@makecaption#1#2{
  \vskip\abovecaptionskip
  \sbox\@tempboxa{#1. #2}
  \ifdim \wd\@tempboxa >\hsize
    #1. #2\par
  \else
    \global \@minipagefalse
    \hb@xt@\hsize{\hfil\box\@tempboxa\hfil}
  \fi
  \vskip\belowcaptionskip}

\makeatother

\def\Ntr{1410}

\def\Nevr{529}

\begin{document}

{\center
{\Large\bf
Tables of rovibronic term values\\
for singlet electronic states of $^{11}$B$^1$H molecule

}
\bigskip

\large
B.~P.~Lavrov and M.~S.~Ryazanov
\bigskip

{\small\it
Institute of Physics, St.-Petersburg State University, St.-Petersburg, 198904, Russia

e-mail: lavrov@pobox.spbu.ru

}
\bigskip

}
\begingroup
\baselineskip=1.6em

In our previous paper \cite{LR} we proposed a new
method for determination of electronic-vibro-rotational (rovibronic)
term values of
diatomics from experimental data on the wavenumbers of
rovibronic spectral lines.
In contrast to existing techniques, this new one is based on the
Rydberg---Ritz principle (Bohr frequency rule) only:
$$
\nu^{n'v'J'}_{n''v''J''} = T_{n'v'J'} - T_{n''v''J''},
$$
where
$\nu^{n'v'J'}_{n''v''J''}$ are meassured wavenumbers, and
$T_{nvJ}$ are corresponding term values, $n$ indicates electronic state,
$v$ and $J$ --- vibrational and rotational quantum numbers,
upper and lower states being marked by single and double primes
correspondingly.

It is shown that a link between a set of rovibronic term values
and a set of wavenumbers of observed rovibronic spectral lines
appears only when three and more different electronic-vibrational (vibronic)
states are pairwise-connected by radiative transitions.
The method differs from known techniques in several aspects, namely, it:

1) doesn't need any assumptions concerning an internal structure of a molecule;

2) doesn't involve any intermediate parameters such
as molecular constants in the traditional approach;

3) gives an opportunity to use in one-stage optimization procedure
all available experimental data obtained for various band systems,
by various authors, and in various works;

4) provides the opportunity of rational selection in an interactive mode
of the experimental data,
eliminating rough errors, to revise wrong identifications of spectral lines
and to compare various sets of experimental data for mutual consistency;

5) allows independent estimation of experimental errors and analysis of
the shape of error distribution;

6) allows to obtain not only an optimal set of rovibronic term values,
but also the error bars determined only by quantity and quality of existing
experimental data.

The method is based on the minimization of the weighted
mean-square deviation
between observed and calculated (as differences of adjustible term values)
values of rovibronic line wavenumbers:
\begin{equation}
r^2=\sum_{\nu^{n'v'J'}_{n''v''J''}}
 \left[\frac{\nu^{n'v'J'}_{n''v''J''}-(T_{n'v'J'}-T_{n''v''J''})}
            {\sigma^{n'v'J'}_{n''v''J''}}\right]^2, \label{r2}
\end{equation}
where
$\sigma^{n'v'J'}_{n''v''J''}$ are their standard deviations,
and sum is over all existing experimental data.
Due to the linearity of the equations used, the optimization problem
comes to solving a system of linear algebraic equations.

The goal of present work was a study of applicability and main features
of the method within an example of certain molecule, namely $^{11}$B$^1$H
isotopomer of boron hydride. This molecule was chosen taking into account
following considerations:

1. BH has the developed rotational structure.

2. There are available spectra
from far infrared (2000~cm$^{-1}$, rotational-vibrational transitions)
to far ultraviolet (70000~cm$^{-1}$, rovibronic transitions),
obtained in many works for many band systems.

3. Non-equilibrium low-pressure plasmas containing boron and hydrogen
species are of interest in astrophysics and numerous applications like
surface treatment of hard metals, production of semiconductors
and thermopolymers, jets of rocket engines and others.
Emission bands of BH molecule are known to be an obligatory spectral feature
of such plasma.

As sources of experimental data on the wavenumbers of singlet rovibronic
spectral lines were taken all know by now to authors works
\cite{Lochte,Thunberg,Almy,Douglas,Bauer,Johns,Lepard,Pianalto,Fernando,Clark}
on this topic.
We restricted ourselves to singlet states and transitions only due to
their simplicity and a circumstance that there is only one paper \cite{Brazier}
on the experimental study of multiplet transitions ($b^3\Sigma^-{-}a^3\Pi$
band system) with partialy unresolved triplet structure.

Grothrian diagram of vibronic levels and the transitions between them
is shown in fig.~1.
In tab.~1 are cited volumes of experimental wavenumbers for each
vibronic band.
One may see that the distribution of investigated lines over vibronic states is
highly nonuniform. For example, the information on rotational levels of ground
vibronic state $X^1\Sigma^+$, $v=0$ contains in wavenumbers of 14 bands
belonging to 11 band systems, explored in 9 various works. At the same time
some of high vibronic and electronic states have only one band
experimentally studied in only one work.
The $K^1\Sigma^+$ state was observed in
\cite{Bauer} only, and identification for rotational lines cannot be
assumed as reliable, thus we did not include this state in present work.

The method under consideration requires mean-square errors of
wavenumbers being used in process of term values determination
for three purposes.
Firstly, weightening of input data in \eqref{r2} depends on these values.
Secondly, selection of input data is based on comparison of deviations
between calculated from derivable term values and meassured wavenumbers
and estimated experimental errors.
Thirdly, the estimations of output term values errors are calculated from
the input wavenumber errors.

Unfortunately only few original works with experimental data on wavenumbers
contain information on accuracy of listed values. And even if error
estimations are mentioned, they seem to be doubtful.
For example, in \cite{Bauer} is stated following:
``The relative accuracy of meassurement of the strongest unblended lines is
thought to be of the order of $\pm0.05$ cm$^{-1}$.'' While there is
absolutely incomprehensible which of the lines are strongest,
what is the accuracy for other lines, etc. In several works only
used spectral resolution is refered.

According to such situation it was necessary to obtain independent
estimations of wavenumber errors, fortunately our method provides
this opportunity.

Ultimate results of these estimations are listed in tab.~2.
During the analysis we detected that data in \cite{Lochte} and \cite{Almy}
contain systematic errors of order $0.05\ldots0.10$ cm$^{-1}$
thus these work were excluded from furter analysis, and results for
them are not shown in the table.
Other work are rather in mutual agreement, except for
$G^1\Pi{-}X^1\Sigma^+$ $(0-0)$ and $H^1\Delta{-}X^1\Sigma^+$ $(0-0)$
bands in \cite{Bauer} and \cite{Lepard}.
These data are systematicly different but we cannot choose one of these
works and reject the other because none of them have undoubted advantage
and no more information on these bands is available.

From remained \Ntr\ wavenumber 32 were rejected as rough errors according
to ``$3\sigma$-rule''.
As a result we found \Nevr\ rovibronic term values for 12 electronic states
of $^{11}$B$^1$H isotopomer of boron hydride molecule.
The values with their standard deviations are listed in tab.~3.

\input{bib.inc}
\endgroup
\newpage
\input{diagram.inc}

\newpage
\input{src.inc}

\newpage
\input{bands.inc}

\input{tab_all.inc}

\end{document}

%% file: diagram.inc
{\center
\includegraphics{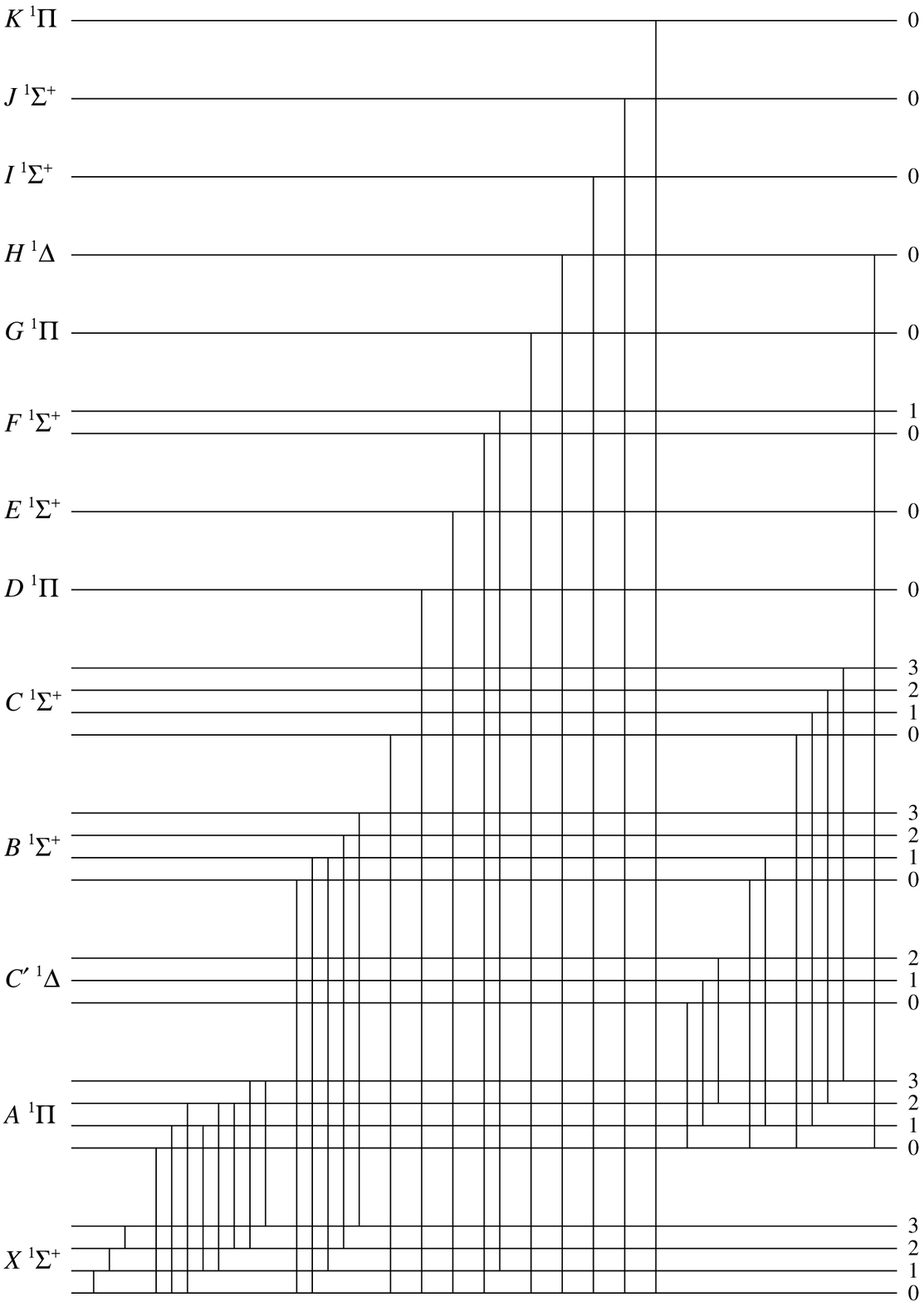}

}
\bigskip
\noindent
\textbf{Figure~1.}\label{diagram}
Grotrian diagram of singlet electronic-vibrational states and
emission and absorption bands experimentaly studied by now.
Symbols of electronic states (in Herzberg notation) are shown
on the left-hand side and vibrational quantum numbers
on the right-hand side of vibronic levels.

%% file: src.inc
\noindent
\textbf{Table~1.} 
Maximum values of rotation quantum number $J''$
of the lower vibronic state for experimentaly studied bands.
Upper indices denote references to original works.
$$
\def\a{\!^\text{\cite{Almy}}}
\def\b{\!^\text{\cite{Bauer}}}
\def\c{\!^\text{\cite{Clark}}}
\def\d{\!^\text{\cite{Douglas}}}
\def\f{\!^\text{\cite{Fernando}}}
\def\h{\!^\text{\cite{Lochte}}}
\def\j{\!^\text{\cite{Johns}}}
\def\l{\!^\text{\cite{Lepard}}}
\def\p{\!^\text{\cite{Pianalto}}}
\def\t{\!^\text{\cite{Thunberg}}}
\begin{array}{r@{\,}l|c|c|c|c|c}
\hline
^1\Lambda'\,- & ^1\Lambda''
   & v'' & v'=0		& v'=1		& v'=2		& v'=3	\\
\hline
X^1\Sigma^+\,- & X^1\Sigma^+
   & 0 & -		& 9\p		& -		& -	\\
 & & 1 & -		& -		& 7\p		& -	\\
 & & 2 & -		& -		& -		& 7\p	\\

A^1\Pi\,- & X^1\Sigma^+
   & 0 & 26\f, 27\j,
	 26\l, 22\t	& 16\j 		& 8\c		& -	\\
 & & 1 & -		& 20\f, 21\j,
 			  20\l, 17\t	& 13\j 		& -	\\
 & & 2 & -		& -		& 14\a, 13\f,
 					  13\j 	& 6\j	\\
 & & 3 & -		& -		& -		& 7\j	\\

B^1\Sigma^+\,- & X^1\Sigma^+
   & 0 & 21\b		& 6\b		& -		& -	\\
 & & 1 & -		& 8\b		& -		& -	\\
 & & 2 & -		& -		& 6\b		& -	\\
 & & 3 & -		& -		& -		& 4\b	\\

C^1\Sigma^+\,- & X^1\Sigma^+
   & 0 & 5\b		& -		& -		& -	\\

D^1\Pi\,- & X^1\Sigma^+
   & 0 & 7\b		& -		& -		& -	\\

E^1\Sigma^+\,- & X^1\Sigma^+
   & 0 & 7\b		& -		& -		& -	\\

F^1\Sigma^+\,- & X^1\Sigma^+
   & 0 & 8\b, 26\l	& -		& -		& -	\\
 & & 1 & -		& 7\b		& -		& -	\\

G^1\Pi\,- & X^1\Sigma^+
 & 0 & 7\b, 14\l	& -		& -		& -	\\

H^1\Delta\,- &  X^1\Sigma^+
 & 0 & 8\b, 15\l	& -		& -		& -	\\

I^1\Sigma^+\,- & X^1\Sigma^+
 & 0 & 6\b		& -		& -		& -	\\

J^1\Sigma^+\,- & X^1\Sigma^+
 & 0 & 3\b		& -		& -		& -	\\

C'^1\Delta\,- & A^1\Pi
   & 0 & 22\j		& -		& -		& -	\\
 & & 1 & -		& 21\j		& -		& -	\\
 & & 2 & -		& -		& 13\j		& -	\\

B^1\Sigma^+\,- & A^1\Pi
   & 0 & 13\d		& -		& -		& -	\\
 & & 1 & -		& 13\d 		& -		& -	\\

C^1\Sigma^+\,- & A^1\Pi
   & 0 & 10\d, 25\j	& -		& -		& -	\\
 & & 1 & -		& 20\j		& -		& -	\\
 & & 2 & -		& -		& 9\j		& -	\\
 & & 3 & -		& -		& -		& 7\j	\\

H^1\Delta\,- & A^1\Pi
   & 0 & 6\b		& -		& -		& -	\\
\hline
\end{array}
$$

%% file: bands.inc
\noindent
\textbf{Table~2.} 
Characteristic of used data by bands.
$\sigma$ --- our estimation for standard deviation of wavenumbers,
$\sigma_\xi$ and $\delta_\xi$ --- obtained standard deviation and bias
of $\xi$ variable (see text).
$$
\def\t{&\multicolumn{2}{c|}{}}
\def\-#1 {\multicolumn{1}{c#1}{-}}
\def\={\phantom{+}}
\arraycolsep=10pt
\begin{array}{c|r@{\,-\,}l|c|r|r|l|c|l}
\hline
\text{Source}
& ^1\Lambda & ^1\Lambda' & v-v' & J'_\text{min} & J'_\text{max} & \sigma, \text{cm}^{-1} & \sigma_\xi & \multicolumn{1}{c}{\delta_\xi} \\
\hline

\text{\cite{Lochte}}
& A^1\Pi & X^1\Sigma^+		& 0-0	& 0	& 26	& \-|	& -	& \- \\
\t				& 1-1	& 0	& 20	& \-|	& -	& \- \\

\text{\cite{Thunberg}}
& A^1\Pi & X^1\Sigma^+		& 0-0	& 0	& 22	& 0.04	& 1.01	& -0.46 \\
\t				& 1-1	& 1	& 17	& 0.05	& 1.05	& +0.004 \\

\text{\cite{Almy}}
& A^1\Pi & X^1\Sigma^+		& 2-2	& 0	& 14	& \-|	& -	& \- \\

\text{\cite{Douglas}}
& B^1\Sigma^+ & A^1\Pi		& 0-0	& 1	& 13	& 0.022	& 1.04	& +0.25 \\
\t				& 1-1	& 1	& 13	& 0.025	& 0.98	& +0.13 \\
& C^1\Sigma^+ & A^1\Pi		& 0-0	& 1	& 10	& 0.064	& 1.02	& +0.37 \\

\text{\cite{Bauer}}
& B^1\Sigma^+ & X^1\Sigma^+	& 0-0	& 0	& 21	& 0.075	& 1.01	& -0.48 \\
\t				& 1-0	& 0	& 6	& 0.12	& 0.99	& -0.06 \\
\t				& 1-1	& 0	& 8	& 0.11	& 0.96	& -0.53 \\
\t				& 2-2	& 0	& 6	& 0.12	& 1.02	& \=0 \\
\t				& 3-3	& 1	& 4	& 0.11	& 1.04	& \=0 \\
& C^1\Sigma^+ & X^1\Sigma^+	& 0-0	& 0	& 5	& 0.088	& 0.93	& -0.64 \\
& D^1\Pi & X^1\Sigma^+		& 0-0	& 0	& 7	& 0.20	& 1.00	& \=0 \\
& E^1\Sigma^+ & X^1\Sigma^+	& 0-0	& 0	& 7	& 0.056	& 0.99	& \=0 \\
& F^1\Sigma^+ & X^1\Sigma^+	& 0-0	& 0	& 8	& 0.14	& 1.05	& -0.33 \\
\t				& 1-1	& 0	& 7	& 0.21	& 1.08	& \=0 \\
& G^1\Pi & X^1\Sigma^+		& 0-0	& 0	& 7	& 0.13	& 1.00	& -0.91 \\
& H^1\Delta & X^1\Sigma^+	& 0-0	& 1	& 8	& 0.11	& 1.01	& -0.77 \\
& I^1\Sigma^+ & X^1\Sigma^+	& 0-0	& 0	& 6	& 0.21	& 1.02	& \=0 \\
& J^1\Sigma^+ & X^1\Sigma^+	& 0-0	& 0	& 3	& 0.28	& 1.00	& \=0 \\
& H^1\Delta & A^1\Pi		& 0-0	& 1	& 6	& 0.26	& 1.05	& -0.01 \\

\text{\cite{Johns}}
& A^1\Pi & X^1\Sigma^+		& 0-0	& 0	& 27	& 0.023	& 0.98	& -0.36 \\
\t				& 1-0	& 0	& 16	& 0.011	& 0.90	& -0.32 \\
\t				& 1-1	& 0	& 21	& 0.026	& 1.03	& -0.09 \\
\t				& 2-1	& 0	& 13	& 0.022	& 1.01	& -0.44 \\
\t				& 2-2	& 0	& 13	& 0.014	& 0.97	& +0.37 \\
\t				& 3-2	& 0	& 6	& 0.06	& 1.00	& -0.44 \\
\t				& 3-3	& 0	& 7	& 0.10	& 0.99	& +0.56 \\
& C^1\Sigma^+ & A^1\Pi		& 0-0	& 1	& 25	& 0.037	& 1.05	& -0.05 \\
\t				& 1-1	& 1	& 20	& 0.034	& 1.01	& -0.005 \\
\t				& 2-2	& 1	& 9	& 0.026	& 1.06	& +0.002 \\
\t				& 3-3	& 1	& 7	& 0.048	& 1.07	& -0.004 \\
& C'^1\Delta & A^1\Pi		& 0-0	& 1	& 22	& 0.021	& 0.92	& +0.01 \\
\t				& 1-1	& 1	& 21	& 0.021	& 1.03	& -0.03 \\
\t				& 2-2	& 1	& 13	& 0.027	& 1.06	& -0.03 \\

\text{\cite{Lepard}}
& F^1\Sigma^+ & X^1\Sigma^+	& 0-0	& 0	& 26	& 0.10	& 1.07	& +0.17 \\
& G^1\Pi & X^1\Sigma^+		& 0-0	& 0	& 14	& 0.07	& 0.98	& +0.87 \\
& H^1\Delta & X^1\Sigma^+	& 0-0	& 2	& 15	& 0.067	& 1.05	& +0.87 \\

\text{\cite{Pianalto}}
& X^1\Sigma^+ & X^1\Sigma^+	& 1-0	& 0	& 9	& 0.0015& 1.02	& +0.25 \\
\t				& 2-1	& 0	& 7	& 0.0015& 0.93	& +0.20 \\
\t				& 3-2	& 2	& 7	& 0.008	& 1.02	& +0.11 \\

\text{\cite{Fernando}}
& A^1\Pi & X^1\Sigma^+		& 0-0	& 0	& 26	& 0.01	& 0.95	& +0.45 \\
\t				& 1-1	& 0	& 20	& 0.013	& 0.97	& +0.34 \\
\t				& 2-2	& 0	& 13	& 0.011	& 0.98	& -0.02 \\

\text{\cite{Clark}}
& A^1\Pi & X^1\Sigma^+		& 2-0	& 0	& 8	& 0.12	& 1.05	& -0.09 \\

\hline
\end{array}
$$

%% file: tab_all.inc
\noindent
\textbf{Table~3.}
Energy levels $E_{n,v,J}$ (with standatd deviations)
of $^{11}$B$^1$H isotopomer of boron hydride in cm$^{-1}$.
$N_\nu$ denotes number of used spectral lines.
$$
\begin{array}{r|r@{}c@{}l|r|r@{}c@{}l|r|r@{}c@{}l|r|r@{}c@{}l|r}
\hline
\multicolumn{17}{c}{X^1\Sigma^+} \\
\hline
\multicolumn{1}{c|}{\raisebox{-1.5ex}[0pt][0pt]{$J$}} &
\multicolumn{4}{c|}{v = 0} &
\multicolumn{4}{c|}{v = 1} &
\multicolumn{4}{c|}{v = 2} &
\multicolumn{4}{c}{v = 3} \\
\cline{2-17}
&
\multicolumn{3}{c|}{E_{n,v,J}} & \multicolumn{1}{c|}{N_\nu} &
\multicolumn{3}{c|}{E_{n,v,J}} & \multicolumn{1}{c|}{N_\nu} &
\multicolumn{3}{c|}{E_{n,v,J}} & \multicolumn{1}{c|}{N_\nu} &
\multicolumn{3}{c|}{E_{n,v,J}} & \multicolumn{1}{c}{N_\nu} \\
\hline
 0 & 0&& & 17            & 2269&.&223(4) & 7    & 4443&.&035(2) & 5    & 6523&.&58(11) & 1 \\
 1 & 23&.&624(4) & 30    & 2292&.&0223(15) & 15 & 4465&.&011(4) & 9    & 6544&.&72(7) & 3  \\
 2 & 70&.&8479(19) & 41  & 2337&.&587(3) & 17   & 4508&.&9455(19) & 13 & 6587&.&108(9) & 6 \\
 3 & 141&.&612(4) & 43   & 2405&.&863(2) & 20   & 4574&.&774(4) & 15   & 6650&.&522(6) & 7 \\
 4 & 235&.&824(2) & 40   & 2496&.&762(3) & 20   & 4662&.&414(2) & 14   & 6734&.&946(7) & 6 \\
 5 & 353&.&372(3) & 38   & 2610&.&172(2) & 20   & 4771&.&752(3) & 15   & 6840&.&276(6) & 5 \\
 6 & 494&.&107(3) & 33   & 2745&.&949(3) & 19   & 4902&.&649(3) & 12   & 6966&.&375(7) & 4 \\
 7 & 657&.&863(4) & 30   & 2903&.&921(3) & 19   & 5054&.&939(4) & 9    & 7113&.&077(8) & 2 \\
 8 & 844&.&427(3) & 26   & 3083&.&899(4) & 15   & 5228&.&421(3) & 7    & 7280&.&188(9) & 1 \\
 9 & 1053&.&583(4) & 20  & 3285&.&647(3) & 13   & 5422&.&895(9) & 6    &&&& \\
10 & 1285&.&059(7) & 20  & 3508&.&920(4) & 13   & 5638&.&087(9) & 6    &&&& \\
11 & 1538&.&591(9) & 17  & 3753&.&427(9) & 12   & 5873&.&751(11) & 6   &&&& \\
12 & 1813&.&840(10) & 18 & 4018&.&898(10) & 11  & 6129&.&586(12) & 6   &&&& \\
13 & 2110&.&527(12) & 16 & 4304&.&990(11) & 11  & 6405&.&226(14) & 4   &&&& \\
14 & 2428&.&240(13) & 16 & 4611&.&333(14) & 9   &&&& &&&& \\
15 & 2766&.&659(15) & 14 & 4937&.&608(15) & 8   &&&& &&&& \\
16 & 3125&.&316(16) & 13 & 5283&.&316(17) & 8   &&&& &&&& \\
17 & 3503&.&842(19) & 10 & 5648&.&21(2) & 8     &&&& &&&& \\
18 & 3901&.&74(2) & 12   & 6031&.&68(4) & 6     &&&& &&&& \\
19 & 4318&.&64(2) & 12   & 6433&.&44(3) & 4     &&&& &&&& \\
20 & 4753&.&98(2) & 12   & 6852&.&90(4) & 3     &&&& &&&& \\
21 & 5207&.&30(3) & 12   & 7289&.&58(4) & 1     &&&& &&&& \\
22 & 5678&.&05(3) & 8    &&&& &&&& &&&& \\
23 & 6165&.&75(3) & 7    &&&& &&&& &&&& \\
24 & 6669&.&72(4) & 5    &&&& &&&& &&&& \\
25 & 7189&.&73(3) & 6    &&&& &&&& &&&& \\
26 & 7724&.&80(4) & 4    &&&& &&&& &&&& \\
27 & 8274&.&64(5) & 1    &&&& &&&& &&&& \\
\hline
\end{array}
$$

\noindent
\textbf{Table~3. (continued)}
$$
\begin{array}{r|r@{}c@{}l|r|r@{}c@{}l|r|r@{}c@{}l|r|r@{}c@{}l|r}
\hline
\multicolumn{17}{c}{A^1\Pi^+} \\
\hline
\multicolumn{1}{c|}{\raisebox{-1.5ex}[0pt][0pt]{$J$}} &
\multicolumn{4}{c|}{v = 0} &
\multicolumn{4}{c|}{v = 1} &
\multicolumn{4}{c|}{v = 2} &
\multicolumn{4}{c}{v = 3} \\
\cline{2-17}
&
\multicolumn{3}{c|}{E_{n,v,J}} & \multicolumn{1}{c|}{N_\nu} &
\multicolumn{3}{c|}{E_{n,v,J}} & \multicolumn{1}{c|}{N_\nu} &
\multicolumn{3}{c|}{E_{n,v,J}} & \multicolumn{1}{c|}{N_\nu} &
\multicolumn{3}{c|}{E_{n,v,J}} & \multicolumn{1}{c}{N_\nu} \\
\hline
 1 & 23097&.&810(6) & 12  & 25183&.&216(6) & 10  & 27012&.&704(7) & 10 & 28490&.&98(3) & 5  \\
 2 & 23145&.&469(7) & 13  & 25227&.&905(6) & 12  & 27053&.&620(6) & 12 & 28526&.&02(3) & 6  \\
 3 & 23216&.&863(6) & 15  & 25294&.&849(6) & 15  & 27114&.&853(6) & 13 & 28578&.&38(3) & 5  \\
 4 & 23311&.&884(6) & 16  & 25383&.&918(6) & 14  & 27196&.&279(6) & 12 & 28647&.&61(3) & 5  \\
 5 & 23430&.&411(6) & 14  & 25494&.&965(6) & 14  & 27297&.&696(6) & 12 & 28733&.&63(5) & 3  \\
 6 & 23572&.&268(6) & 15  & 25627&.&796(6) & 15  & 27418&.&844(6) & 12 & 28835&.&61(7) & 3  \\
 7 & 23737&.&235(6) & 14  & 25782&.&184(6) & 14  & 27559&.&448(7) & 12 & 28952&.&84(12) & 1 \\
 8 & 23925&.&089(7) & 13  & 25957&.&833(6) & 12  & 27719&.&129(7) & 11 &&&& \\
 9 & 24135&.&526(7) & 14  & 26154&.&459(6) & 12  & 27897&.&451(7) & 10 &&&& \\
10 & 24368&.&265(8) & 12  & 26371&.&677(7) & 13  & 28093&.&963(10) & 8 &&&& \\
11 & 24622&.&941(9) & 11  & 26609&.&097(8) & 12  & 28308&.&053(10) & 8 &&&& \\
12 & 24899&.&204(11) & 12 & 26866&.&273(10) & 13 & 28539&.&098(12) & 7 &&&& \\
13 & 25196&.&598(12) & 11 & 27142&.&700(11) & 13 & 28786&.&382(16) & 3 &&&& \\
14 & 25514&.&715(13) & 11 & 27437&.&894(13) & 11 &&&& &&&& \\
15 & 25853&.&052(15) & 11 & 27751&.&143(16) & 9  &&&& &&&& \\
16 & 26211&.&113(17) & 10 & 28081&.&941(19) & 8  &&&& &&&& \\
17 & 26588&.&298(18) & 9  & 28429&.&39(3) & 7    &&&& &&&& \\
18 & 26984&.&08(2) & 8    & 28792&.&94(2) & 5    &&&& &&&& \\
19 & 27397&.&76(2) & 9    & 29171&.&49(4) & 5    &&&& &&&& \\
20 & 27828&.&74(2) & 8    & 29564&.&05(4) & 4    &&&& &&&& \\
21 & 28276&.&23(3) & 8 &&&& &&&& &&&& \\
22 & 28739&.&49(3) & 8 &&&& &&&& &&&& \\
23 & 29217&.&67(3) & 3 &&&& &&&& &&&& \\
24 & 29709&.&92(3) & 4 &&&& &&&& &&&& \\
25 & 30215&.&20(4) & 3 &&&& &&&& &&&& \\
26 & 30732&.&46(4) & 2 &&&& &&&& &&&& \\
\hline
\end{array}
$$

\noindent
\textbf{Table~3. (continued)}
$$
\begin{array}{r|r@{}c@{}l|r|r@{}c@{}l|r|r@{}c@{}l|r|r@{}c@{}l|r}
\hline
\multicolumn{17}{c}{A^1\Pi^-} \\
\hline
\multicolumn{1}{c|}{\raisebox{-1.5ex}[0pt][0pt]{$J$}} &
\multicolumn{4}{c|}{v = 0} &
\multicolumn{4}{c|}{v = 1} &
\multicolumn{4}{c|}{v = 2} &
\multicolumn{4}{c}{v = 3} \\
\cline{2-17}
&
\multicolumn{3}{c|}{E_{n,v,J}} & \multicolumn{1}{c|}{N_\nu} &
\multicolumn{3}{c|}{E_{n,v,J}} & \multicolumn{1}{c|}{N_\nu} &
\multicolumn{3}{c|}{E_{n,v,J}} & \multicolumn{1}{c|}{N_\nu} &
\multicolumn{3}{c|}{E_{n,v,J}} & \multicolumn{1}{c}{N_\nu} \\
\hline
 1 & 23097&.&736(9) & 7  & 25183&.&137(8) & 7  & 27012&.&618(10) & 6 & 28490&.&95(4) & 3  \\
 2 & 23145&.&257(8) & 8  & 25227&.&701(8) & 8  & 27053&.&441(8) & 7  & 28525&.&90(4) & 3  \\
 3 & 23216&.&408(8) & 9  & 25294&.&427(7) & 9  & 27114&.&478(8) & 8  & 28578&.&05(4) & 3  \\
 4 & 23311&.&139(8) & 8  & 25383&.&235(7) & 9  & 27195&.&678(8) & 8  & 28647&.&22(5) & 3  \\
 5 & 23429&.&297(8) & 10 & 25493&.&939(7) & 9  & 27296&.&790(8) & 8  & 28732&.&93(5) & 3  \\
 6 & 23570&.&718(8) & 8  & 25626&.&365(7) & 9  & 27417&.&600(8) & 8  & 28834&.&62(10) & 3 \\
 7 & 23735&.&205(8) & 9  & 25780&.&295(7) & 8  & 27557&.&798(8) & 8  & 28951&.&8(2) & 1   \\
 8 & 23922&.&468(8) & 8  & 25955&.&424(7) & 9  & 27717&.&021(8) & 8  &&&& \\
 9 & 24132&.&286(9) & 9  & 26151&.&498(8) & 9  & 27894&.&869(11) & 5 &&&& \\
10 & 24364&.&350(10) & 9 & 26368&.&090(8) & 9  & 28090&.&832(11) & 5 &&&& \\
11 & 24618&.&301(11) & 7 & 26604&.&866(10) & 9 & 28304&.&392(13) & 5 &&&& \\
12 & 24893&.&772(13) & 7 & 26861&.&325(11) & 9 & 28534&.&855(14) & 4 &&&& \\
13 & 25190&.&366(14) & 6 & 27137&.&057(13) & 7 & 28781&.&445(15) & 3 &&&& \\
14 & 25507&.&613(16) & 4 & 27431&.&438(14) & 8 &&&& &&&& \\
15 & 25845&.&069(17) & 4 & 27743&.&981(16) & 6 &&&& &&&& \\
16 & 26202&.&187(18) & 5 & 28073&.&912(17) & 6 &&&& &&&& \\
17 & 26578&.&48(2) & 4   & 28420&.&69(2) & 5   &&&& &&&& \\
18 & 26973&.&26(2) & 5   & 28783&.&34(4) & 4   &&&& &&&& \\
19 & 27386&.&02(2) & 5   & 29161&.&19(3) & 4   &&&& &&&& \\
20 & 27816&.&01(3) & 5   & 29552&.&94(5) & 1   &&&& &&&& \\
21 & 28262&.&58(3) & 5 &&&& &&&& &&&& \\
22 & 28724&.&89(3) & 4 &&&& &&&& &&&& \\
23 & 29202&.&16(3) & 3 &&&& &&&& &&&& \\
24 & 29693&.&44(4) & 3 &&&& &&&& &&&& \\
25 & 30198&.&00(3) & 3 &&&& &&&& &&&& \\
26 & 30714&.&41(5) & 1 &&&& &&&& &&&& \\
\hline
\end{array}
$$

\noindent
\textbf{Table~3. (continued)}
$$
\begin{array}{r|r@{}c@{}l|r|r@{}c@{}l|r|r@{}c@{}l|r}
\hline
\multicolumn{13}{c}{C'^1\Delta^+} \\
\hline
\multicolumn{1}{c|}{\raisebox{-1.5ex}[0pt][0pt]{$J$}} &
\multicolumn{4}{c|}{v = 0} &
\multicolumn{4}{c|}{v = 1} &
\multicolumn{4}{c}{v = 2} \\
\cline{2-13}
&
\multicolumn{3}{c|}{E_{n,v,J}} & \multicolumn{1}{c|}{N_\nu} &
\multicolumn{3}{c|}{E_{n,v,J}} & \multicolumn{1}{c|}{N_\nu} &
\multicolumn{3}{c|}{E_{n,v,J}} & \multicolumn{1}{c}{N_\nu} \\
\hline
 2 & 46178&.&53(2) & 3   & 48694&.&490(15) & 3 & 51117&.&272(19) & 3 \\
 3 & 46253&.&864(13) & 3 & 48767&.&452(13) & 3 & 51187&.&915(16) & 3 \\
 4 & 46354&.&126(13) & 3 & 48864&.&583(13) & 3 & 51281&.&968(19) & 3 \\
 5 & 46479&.&248(13) & 3 & 48985&.&794(13) & 3 & 51399&.&32(2) & 3   \\
 6 & 46629&.&091(13) & 3 & 49130&.&937(13) & 3 & 51539&.&850(19) & 3 \\
 7 & 46803&.&465(16) & 3 & 49299&.&860(13) & 3 & 51703&.&37(2) & 3   \\
 8 & 47002&.&155(16) & 2 & 49492&.&299(13) & 3 & 51889&.&695(16) & 3 \\
 9 & 47224&.&979(16) & 3 & 49708&.&190(13) & 3 & 52098&.&64(2) & 3   \\
10 & 47471&.&651(14) & 3 & 49947&.&110(16) & 2 & 52329&.&82(2) & 2   \\
11 & 47741&.&906(15) & 3 & 50208&.&872(16) & 3 & 52583&.&20(3) & 1   \\
12 & 48035&.&445(15) & 3 & 50493&.&160(15) & 3 & 52858&.&30(11) & 1  \\
13 & 48351&.&932(17) & 3 & 50799&.&604(19) & 3 &&&& \\
14 & 48691&.&036(19) & 2 & 51127&.&920(19) & 2 &&&& \\
15 & 49052&.&35(2) & 1   & 51477&.&81(2) & 2   &&&& \\
16 & 49435&.&44(2) & 2   & 51847&.&85(9) & 1   &&&& \\
17 & 49840&.&09(14) & 2  & 52240&.&35(3) & 1   &&&& \\
18 & 50265&.&25(9) & 1   & 52652&.&22(4) & 1   &&&& \\
19 & 50711&.&56(3) & 1   & 53083&.&57(9) & 1   &&&& \\
20 & 51177&.&71(3) & 1 &&&& &&&& \\
21 & 51663&.&35(3) & 1 &&&& &&&& \\
\hline
\end{array}
$$
$$
\begin{array}{r|r@{}c@{}l|r|r@{}c@{}l|r|r@{}c@{}l|r}
\hline
\multicolumn{13}{c}{C'^1\Delta^-} \\
\hline
\multicolumn{1}{c|}{\raisebox{-1.5ex}[0pt][0pt]{$J$}} &
\multicolumn{4}{c|}{v = 0} &
\multicolumn{4}{c|}{v = 1} &
\multicolumn{4}{c}{v = 2} \\
\cline{2-13}
&
\multicolumn{3}{c|}{E_{n,v,J}} & \multicolumn{1}{c|}{N_\nu} &
\multicolumn{3}{c|}{E_{n,v,J}} & \multicolumn{1}{c|}{N_\nu} &
\multicolumn{3}{c|}{E_{n,v,J}} & \multicolumn{1}{c}{N_\nu} \\
\hline
 2 & 46178&.&54(2) & 3   & 48694&.&484(15) & 3 & 51117&.&24(3) & 3   \\
 3 & 46253&.&846(13) & 3 & 48767&.&438(15) & 3 & 51187&.&906(19) & 3 \\
 4 & 46354&.&121(13) & 3 & 48864&.&573(13) & 3 & 51281&.&977(16) & 3 \\
 5 & 46479&.&247(13) & 3 & 48985&.&809(13) & 3 & 51399&.&320(19) & 3 \\
 6 & 46629&.&096(13) & 3 & 49130&.&944(13) & 3 & 51539&.&850(16) & 3 \\
 7 & 46803&.&450(13) & 3 & 49299&.&858(13) & 3 & 51703&.&367(16) & 3 \\
 8 & 47002&.&24(14) & 2  & 49492&.&357(16) & 2 & 51889&.&698(17) & 3 \\
 9 & 47224&.&978(14) & 3 & 49708&.&170(16) & 2 & 52098&.&613(17) & 3 \\
10 & 47471&.&668(14) & 3 & 49947&.&118(14) & 3 & 52329&.&86(2) & 2   \\
11 & 47741&.&913(17) & 2 & 50208&.&865(14) & 3 & 52583&.&21(3) & 1   \\
12 & 48035&.&452(18) & 2 & 50493&.&146(17) & 2 & 52858&.&35(3) & 1   \\
13 & 48351&.&924(19) & 2 & 50799&.&602(16) & 3 & 53155&.&02(3) & 1   \\
14 & 48691&.&00(2) & 2   & 51127&.&981(19) & 2 &&&& \\
15 & 49052&.&33(2) & 2   & 51477&.&78(2) & 2   &&&& \\
16 & 49435&.&46(2) & 2   & 51848&.&71(3) & 1   &&&& \\
17 & 49839&.&98(2) & 2   & 52240&.&27(4) & 1   &&&& \\
18 & 50265&.&53(3) & 1   & 52652&.&23(3) & 1   &&&& \\
19 & 50711&.&56(3) & 1   & 53083&.&99(4) & 1   &&&& \\
20 & 51177&.&67(3) & 1   & 53535&.&01(4) & 1   &&&& \\
21 & 51663&.&35(3) & 1 &&&& &&&& \\
22 & 52168&.&12(3) & 1 &&&& &&&& \\
\hline
\end{array}
$$

\noindent
\textbf{Table~3. (continued)}
$$
\begin{array}{r|r@{}c@{}l|r|r@{}c@{}l|r|r@{}c@{}l|r|r@{}c@{}l|r}
\hline
\multicolumn{17}{c}{B^1\Sigma^+} \\
\hline
\multicolumn{1}{c|}{\raisebox{-1.5ex}[0pt][0pt]{$J$}} &
\multicolumn{4}{c|}{v = 0} &
\multicolumn{4}{c|}{v = 1} &
\multicolumn{4}{c|}{v = 2} &
\multicolumn{4}{c}{v = 3} \\
\cline{2-17}
&
\multicolumn{3}{c|}{E_{n,v,J}} & \multicolumn{1}{c|}{N_\nu} &
\multicolumn{3}{c|}{E_{n,v,J}} & \multicolumn{1}{c|}{N_\nu} &
\multicolumn{3}{c|}{E_{n,v,J}} & \multicolumn{1}{c|}{N_\nu} &
\multicolumn{3}{c|}{E_{n,v,J}} & \multicolumn{1}{c}{N_\nu} \\
\hline
 0 & 52346&.&70(2) & 2   & 54594&.&79(2) & 3   & 56690&.&26(8) & 2  & 58564&.&61(11) & 1 \\
 1 & 52370&.&968(17) & 4 & 54617&.&846(18) & 5 & 56733&.&45(8) & 2  & 58605&.&06(8) & 2  \\
 2 & 52419&.&279(13) & 5 & 54663&.&809(15) & 7 & 56798&.&71(8) & 2  & 58665&.&55(8) & 2  \\
 3 & 52491&.&644(13) & 5 & 54732&.&755(15) & 7 & 56885&.&20(8) & 2  & 58746&.&10(11) & 1 \\
 4 & 52587&.&978(13) & 5 & 54824&.&470(15) & 7 & 56993&.&21(8) & 2  &&&& \\
 5 & 52708&.&185(13) & 5 & 54938&.&809(18) & 5 & 57122&.&25(12) & 1 &&&& \\
 6 & 52852&.&061(13) & 5 & 55075&.&779(18) & 5 &&&& &&&& \\
 7 & 53019&.&456(13) & 5 & 55235&.&26(2) & 3   &&&& &&&& \\
 8 & 53210&.&138(13) & 5 & 55416&.&362(18) & 3 &&&& &&&& \\
 9 & 53423&.&879(13) & 5 & 55619&.&563(15) & 4 &&&& &&&& \\
10 & 53660&.&354(14) & 5 & 55844&.&309(19) & 2 &&&& &&&& \\
11 & 53919&.&276(15) & 5 & 56090&.&26(2) & 2   &&&& &&&& \\
12 & 54200&.&32(2) & 2   & 56357&.&08(2) & 2   &&&& &&&& \\
13 & 54503&.&08(2) & 3 &&&& &&&& &&&& \\
14 & 54827&.&17(5) & 2 &&&& &&&& &&&& \\
15 & 55172&.&23(5) & 2 &&&& &&&& &&&& \\
16 & 55537&.&61(6) & 2 &&&& &&&& &&&& \\
17 & 55923&.&46(6) & 2 &&&& &&&& &&&& \\
18 & 56328&.&53(6) & 2 &&&& &&&& &&&& \\
19 & 56752&.&52(6) & 2 &&&& &&&& &&&& \\
20 & 57194&.&86(6) & 2 &&&& &&&& &&&& \\
21 & 57655&.&23(8) & 1 &&&& &&&& &&&& \\
22 & 58132&.&60(8) & 1 &&&& &&&& &&&& \\
\hline
\end{array}
$$

\noindent
\textbf{Table~3. (continued)}
$$
\begin{array}{r|r@{}c@{}l|r|r@{}c@{}l|r|r@{}c@{}l|r|r@{}c@{}l|r}
\hline
\multicolumn{17}{c}{C^1\Sigma^+} \\
\hline
\multicolumn{1}{c|}{\raisebox{-1.5ex}[0pt][0pt]{$J$}} &
\multicolumn{4}{c|}{v = 0} &
\multicolumn{4}{c|}{v = 1} &
\multicolumn{4}{c|}{v = 2} &
\multicolumn{4}{c}{v = 3} \\
\cline{2-17}
&
\multicolumn{3}{c|}{E_{n,v,J}} & \multicolumn{1}{c|}{N_\nu} &
\multicolumn{3}{c|}{E_{n,v,J}} & \multicolumn{1}{c|}{N_\nu} &
\multicolumn{3}{c|}{E_{n,v,J}} & \multicolumn{1}{c|}{N_\nu} &
\multicolumn{3}{c|}{E_{n,v,J}} & \multicolumn{1}{c}{N_\nu} \\
\hline
 0 & 55333&.&72(3) & 2   & 57700&.&26(3) & 1  & 59960&.&28(3) & 1 &&&&  \\
 1 & 55358&.&13(2) & 6   & 57723&.&90(14) & 2 & 59982&.&838(19) & 2 & 62139&.&97(5) & 2  \\
 2 & 55406&.&872(19) & 7 & 57770&.&78(2) & 3  & 60028&.&083(16) & 3 & 62180&.&69(4) & 3  \\
 3 & 55479&.&91(2) & 7   & 57841&.&19(2) & 2  & 60095&.&791(16) & 3 & 62244&.&85(4) & 3  \\
 4 & 55577&.&15(2) & 7   & 57934&.&98(2) & 3  & 60185&.&951(16) & 3 & 62331&.&12(5) & 2  \\
 5 & 55698&.&467(19) & 6 & 58051&.&90(2) & 3  & 60298&.&422(16) & 3 & 62438&.&86(6) & 2  \\
 6 & 55843&.&76(2) & 5   & 58191&.&96(2) & 3  & 60432&.&992(16) & 3 & 62567&.&39(11) & 2 \\
 7 & 56012&.&80(2) & 5   & 58354&.&92(2) & 3  & 60589&.&364(16) & 3 &&&& \\
 8 & 56205&.&38(2) & 5   & 58540&.&59(2) & 2  & 60766&.&650(16) & 3 &&&& \\
 9 & 56421&.&31(2) & 4   & 58748&.&72(2) & 2  &&&& &&&& \\
10 & 56660&.&33(2) & 4   & 58979&.&05(2) & 3  &&&& &&&& \\
11 & 56922&.&07(2) & 3   & 59231&.&32(2) & 3  &&&& &&&& \\
12 & 57206&.&34(2) & 3   & 59505&.&17(2) & 3  &&&& &&&& \\
13 & 57512&.&72(2) & 3   & 59800&.&39(3) & 3  &&&& &&&& \\
14 & 57840&.&89(2) & 3   & 60116&.&47(3) & 3  &&&& &&&& \\
15 & 58190&.&45(3) & 3   & 60453&.&05(3) & 2  &&&& &&&& \\
16 & 58561&.&01(3) & 3   & 60809&.&87(3) & 3  &&&& &&&& \\
17 & 58952&.&11(3) & 3   & 61186&.&39(3) & 3  &&&& &&&& \\
18 & 59363&.&27(3) & 3   & 61582&.&04(4) & 2  &&&& &&&& \\
19 & 59794&.&13(3) & 3   & 61996&.&44(4) & 2  &&&& &&&& \\
20 & 60244&.&07(3) & 3 &&&& &&&& &&&& \\
21 & 60712&.&65(4) & 2 &&&& &&&& &&&& \\
22 & 61199&.&31(4) & 2 &&&& &&&& &&&& \\
23 & 61703&.&63(4) & 2 &&&& &&&& &&&& \\
24 & 62224&.&77(5) & 1 &&&& &&&& &&&& \\
25 & 62762&.&53(5) & 1 &&&& &&&& &&&& \\
\hline
\end{array}
$$

\bigskip

$$
\begin{array}{r|r@{}c@{}l|r}
\hline
\multicolumn{5}{c}{D^1\Pi^+} \\
\hline
\multicolumn{1}{c|}{\raisebox{-1.5ex}[0pt][0pt]{$J$}} &
\multicolumn{4}{c}{v = 0} \\
\cline{2-5}
&
\multicolumn{3}{c|}{E_{n,v,J}} & \multicolumn{1}{c}{N_\nu} \\
\hline
 1 & 61128&.&87(14) & 2 \\
 2 & 61175&.&68(14) & 2 \\
 3 & 61245&.&93(14) & 2 \\
 4 & 61339&.&57(14) & 2 \\
 5 & 61456&.&7(2) & 1 \\
 6 & 61596&.&6(2) & 1 \\
 7 & 61760&.&0(2) & 1 \\
\hline
\end{array}
\qquad
\begin{array}{r|r@{}c@{}l|r}
\hline
\multicolumn{5}{c}{D^1\Pi^-} \\
\hline
\multicolumn{1}{c|}{\raisebox{-1.5ex}[0pt][0pt]{$J$}} &
\multicolumn{4}{c}{v = 0} \\
\cline{2-5}
&
\multicolumn{3}{c|}{E_{n,v,J}} & \multicolumn{1}{c}{N_\nu} \\
\hline
 1 & 61130&.&0(2) & 1 \\
 2 & 61179&.&3(2) & 1 \\
 3 & 61253&.&1(2) & 1 \\
 4 & 61351&.&4(2) & 1 \\
 5 & 61474&.&0(2) & 1 \\
 6 & 61620&.&8(2) & 1 \\
 7 & 61791&.&5(2) & 1 \\
\hline
\end{array}
%
\qquad\qquad
\begin{array}{r|r@{}c@{}l|r}
\hline
\multicolumn{5}{c}{E^1\Sigma^+} \\
\hline
\multicolumn{1}{c|}{\raisebox{-1.5ex}[0pt][0pt]{$J$}} &
\multicolumn{4}{c}{v = 0} \\
\cline{2-5}
&
\multicolumn{3}{c|}{E_{n,v,J}} & \multicolumn{1}{c}{N_\nu} \\
\hline
 0 & 61872&.&32(6) & 1 \\
 1 & 61898&.&53(4) & 2 \\
 2 & 61950&.&64(4) & 2 \\
 3 & 62028&.&83(4) & 2 \\
 4 & 62132&.&63(4) & 2 \\
 5 & 62262&.&12(6) & 1 \\
 6 & 62416&.&87(6) & 1 \\
\hline
\end{array}
$$

\noindent
\textbf{Table~3. (continued)}
$$
\begin{array}{r|r@{}c@{}l|r|r@{}c@{}l|r}
\hline
\multicolumn{9}{c}{F^1\Sigma^+} \\
\hline
\multicolumn{1}{c|}{\raisebox{-1.5ex}[0pt][0pt]{$J$}} &
\multicolumn{4}{c|}{v = 0} &
\multicolumn{4}{c}{v = 1} \\
\cline{2-9}
&
\multicolumn{3}{c|}{E_{n,v,J}} & \multicolumn{1}{c|}{N_\nu} &
\multicolumn{3}{c|}{E_{n,v,J}} & \multicolumn{1}{c}{N_\nu} \\
\hline
 0 & 66078&.&1(14) & 2 & 68268&.&9(2) & 1   \\
 1 & 66090&.&37(6) & 4 & 68285&.&0(4) & 2   \\
 2 & 66119&.&29(6) & 4 & 68316&.&4(2) & 2   \\
 3 & 66165&.&20(7) & 4 & 68366&.&0(2) & 2   \\
 4 & 66229&.&99(6) & 4 & 68433&.&8(2) & 2   \\
 5 & 66315&.&25(6) & 4 & 68520&.&9(2) & 2   \\
 6 & 66421&.&69(7) & 3 & 68628&.&51(15) & 2 \\
 7 & 66550&.&20(6) & 4  &&&& \\
 8 & 66700&.&71(8) & 2  &&&& \\
 9 & 66875&.&03(7) & 2  &&&& \\
10 & 67071&.&30(10) & 1 &&&& \\
11 & 67290&.&00(7) & 2  &&&& \\
12 & 67531&.&20(7) & 2  &&&& \\
13 & 67794&.&50(7) & 2  &&&& \\
14 & 68079&.&73(7) & 2  &&&& \\
15 & 68386&.&93(7) & 2  &&&& \\
16 & 68714&.&95(10) & 1 &&&& \\
17 & 69063&.&37(10) & 1 &&&& \\
18 & 69433&.&2(2) & 1   &&&& \\
19 & 69823&.&84(10) & 1 &&&& \\
20 & 70232&.&81(10) & 1 &&&& \\
21 & 70660&.&92(10) & 1 &&&& \\
22 & 71107&.&50(10) & 1 &&&& \\
23 & 71571&.&95(10) & 1 &&&& \\
24 & 72053&.&05(10) & 1 &&&& \\
25 & 72551&.&57(11) & 1 &&&& \\
26 & 73064&.&11(10) & 1 &&&& \\
27 & 73594&.&11(11) & 1 &&&& \\
\hline
\end{array}
$$

\bigskip

$$
\begin{array}{r|r@{}c@{}l|r}
\hline
\multicolumn{5}{c}{G^1\Pi^+} \\
\hline
\multicolumn{1}{c|}{\raisebox{-1.5ex}[0pt][0pt]{$J$}} &
\multicolumn{4}{c}{v = 0} \\
\cline{2-5}
&
\multicolumn{3}{c|}{E_{n,v,J}} & \multicolumn{1}{c}{N_\nu} \\
\hline
 1 & 66407&.&42(4) & 4 \\
 2 & 66444&.&86(6) & 2 \\
 3 & 66510&.&26(6) & 2 \\
 4 & 66602&.&48(6) & 2 \\
 5 & 66720&.&47(7) & 1 \\
 6 & 66863&.&77(7) & 1 \\
 7 & 67031&.&34(7) & 1 \\
 8 & 67222&.&77(7) & 1 \\
 9 & 67439&.&51(7) & 1 \\
10 & 67677&.&98(7) & 1 \\
11 & 67939&.&84(7) & 1 \\
12 & 68224&.&22(7) & 1 \\
13 & 68530&.&63(7) & 1 \\
\hline
\end{array}
\qquad
\begin{array}{r|r@{}c@{}l|r}
\hline
\multicolumn{5}{c}{G^1\Pi^-} \\
\hline
\multicolumn{1}{c|}{\raisebox{-1.5ex}[0pt][0pt]{$J$}} &
\multicolumn{4}{c}{v = 0} \\
\cline{2-5}
&
\multicolumn{3}{c|}{E_{n,v,J}} & \multicolumn{1}{c}{N_\nu} \\
\hline
 1 & 66398&.&06(18) & 2 \\
 2 & 66421&.&97(6) & 2 \\
 3 & 66471&.&09(6) & 2 \\
 4 & 66544&.&75(6) & 2 \\
 5 & 66642&.&82(6) & 2 \\
 6 & 66765&.&17(6) & 2 \\
 7 & 66911&.&69(6) & 2 \\
 8 & 67082&.&04(7) & 1 \\
 9 & 67276&.&15(7) & 1 \\
10 & 67493&.&63(7) & 1 \\
11 & 67734&.&24(7) & 1 \\
12 & 67997&.&97(7) & 1 \\
13 & 68284&.&09(7) & 1 \\
\hline
\end{array}
$$

\noindent
\textbf{Table~3. (continued)}
$$
\begin{array}[t]{r|r@{}c@{}l|r}
\hline
\multicolumn{5}{c}{H^1\Delta^+} \\
\hline
\multicolumn{1}{c|}{\raisebox{-1.5ex}[0pt][0pt]{$J$}} &
\multicolumn{4}{c}{v = 0} \\
\cline{2-5}
&
\multicolumn{3}{c|}{E_{n,v,J}} & \multicolumn{1}{c}{N_\nu} \\
\hline
 2 & 66548&.&30(5) & 5 \\
 3 & 66654&.&40(5) & 4 \\
 4 & 66784&.&91(5) & 3 \\
 5 & 66946&.&90(6) & 3 \\
 6 & 67129&.&62(7) & 1 \\
 7 & 67336&.&34(7) & 1 \\
 8 & 67566&.&74(7) & 1 \\
 9 & 67819&.&91(7) & 1 \\
10 & 68095&.&33(7) & 1 \\
11 & 68392&.&38(7) & 1 \\
12 & 68710&.&50(7) & 1 \\
13 & 69049&.&08(7) & 1 \\
\hline
\end{array}
\qquad
\begin{array}[t]{r|r@{}c@{}l|r}
\hline
\multicolumn{5}{c}{H^1\Delta^-} \\
\hline
\multicolumn{1}{c|}{\raisebox{-1.5ex}[0pt][0pt]{$J$}} &
\multicolumn{4}{c}{v = 0} \\
\cline{2-5}
&
\multicolumn{3}{c|}{E_{n,v,J}} & \multicolumn{1}{c}{N_\nu} \\
\hline
 2 & 66543&.&04(5) & 4 \\
 3 & 66641&.&07(6) & 2 \\
 4 & 66763&.&06(6) & 4 \\
 5 & 66909&.&05(6) & 2 \\
 6 & 67081&.&07(6) & 2 \\
 7 & 67273&.&54(6) & 2 \\
 8 & 67490&.&70(6) & 2 \\
 9 & 67731&.&02(7) & 1 \\
10 & 67994&.&27(7) & 1 \\
11 & 68280&.&04(7) & 1 \\
12 & 68587&.&84(7) & 1 \\
13 &     -& &      &   \\
14 & 69267&.&99(7) & 1 \\
15 & 69637&.&59(7) & 1 \\
\hline
\end{array}
$$

\bigskip

$$
\begin{array}{r|r@{}c@{}l|r}
\hline
\multicolumn{5}{c}{I^1\Sigma^+} \\
\hline
\multicolumn{1}{c|}{\raisebox{-1.5ex}[0pt][0pt]{$J$}} &
\multicolumn{4}{c}{v = 0} \\
\cline{2-5}
&
\multicolumn{3}{c|}{E_{n,v,J}} & \multicolumn{1}{c}{N_\nu} \\
\hline
 0 & 67395&.&8(2) & 1 \\
 1 & 67420&.&43(15) & 2 \\
 2 & 67469&.&78(15) & 2 \\
 3 & 67543&.&90(15) & 2 \\
 4 & 67642&.&24(15) & 2 \\
 5 & 67765&.&03(15) & 2 \\
 6 & 67912&.&1(2) & 1 \\
\hline
\end{array}
%
\qquad\qquad
\begin{array}{r|r@{}c@{}l|r}
\hline
\multicolumn{5}{c}{J^1\Sigma^+} \\
\hline
\multicolumn{1}{c|}{\raisebox{-1.5ex}[0pt][0pt]{$J$}} &
\multicolumn{4}{c}{v = 0} \\
\cline{2-5}
&
\multicolumn{3}{c|}{E_{n,v,J}} & \multicolumn{1}{c}{N_\nu} \\
\hline
 1 & 70056&.&8(2) & 2 \\
 2 & 70092&.&1(2) & 2 \\
\hline
\end{array}
$$

%% file: BHterms.bbl
\begin{thebibliography}{99}
\topsep=0pt
\partopsep=0pt
\parskip=0pt
\parsep=0pt
\itemsep=0pt

\bibitem{LR}
Lavrov~B.P., Ryazanov~M.S. //
Khimicheskaya Fizika (Soviet Journal of Chemical Physics), in print (2005);
e-print: physics/0405132 at http://arXiv.org.

\bibitem{Lochte}
Lochte-Holtgreven~W., van der Vleugel~E.S. //
ZS. f. Phys. \textbf{70}, 188 (1931).

\bibitem{Thunberg}
Thunberg~S.F. //
ZS. f. Phys. \textbf{100}, 471 (1936).

\bibitem{Almy}
Almy~G.M., Horsfall~R.B., Jr //
Phys. Rev. \textbf{51}, 491 (1937).

\bibitem{Douglas}
Douglas~A.E. //
Canad. J. Res. \textbf{19} A, 27 (1941).

\bibitem{Bauer}
Bauer~S.H., Herzberg~G., Johns~J.W. //
J. Mol. Spectr. \textbf{13}, 256 (1964).

\bibitem{Johns}
Johns~J.W., Grimm~F.A., Porter~R.F. //
J. Mol. Spectr. \textbf{22}, 435 (1967).

\bibitem{Lepard}
Johns~J.W., Lepard~D.W. //
J. Mol. Spectr. \textbf{55}, 374 (1975).

\bibitem{Pianalto}
Pianalto~F.S., O'Brien~L.C., Keller~P.C., Bernath~P.F. //
J. Mol. Spectr. \textbf{129}, 348 (1988).

\bibitem{Fernando}
Fernando~W.T.M.L., Bernath~P.F. //
J. Mol. Spectr. \textbf{145}, 329 (1991).

\bibitem{Clark}
Clark J., Konopka M., Zhang L.-M., Grant E.R. //
Chem. Phys. Lett. \textbf{340}, 45 (2001).

\bibitem{Brazier}
Brazier~C.R. //
J. Mol. Spectr. \textbf{177}, 90 (1996).

\end{thebibliography}
